\documentclass[a4paper, 10pt, conference]{ieeeconf}      

\usepackage[pdftex]{graphicx}
\usepackage{epstopdf}
\usepackage{subfigure}

\IEEEoverridecommandlockouts                              

\usepackage[draft,marginclue,footnote]{fixme}
\FXRegisterAuthor{aa}{anaa}{AA}

\title{\LARGE \bf Data Gathering for Sensing Applications in Vehicular Networks (Poster)}

\author{Mohammad Nozari Zarmehri and Ana Aguiar\\
	Faculdade de Engenharia da Universidade do Porto (FEUP)\\
	Instituto de Telecomunica\c c\~ oes (IT)\\
	\{mohammad.nozari, ana.aguiar\}@fe.up.pt
\thanks{M. Nozari's work is supported by the Funda\c c\~ ao para Ci\^ encia e Tecnologia under the grant SFRH/BD/71438/2010 and by the CMU|Portugal project Drive-In.}
\thanks{A. Aguiar is with the University of Porto and Instituto de Telecomunica\c c\~ oes, Porto, Portugal.}
}

\begin{document}

\makeatletter
\newenvironment{tablehere}
  {\def\@captype{table}}
  {}

\newenvironment{figurehere}
  {\def\@captype{figure}}
  {}
\makeatother

\maketitle
\thispagestyle{empty}
\pagestyle{empty}

\begin{abstract}
We propose to use Vehicular ad hoc networks (VANET) as the infrastructure for an urban cyber-physical system for gathering up-to-date data about a city, like traffic conditions or environmental parameters. In this context, it is critical to design a data collection protocol that enables retrieving the data from the vehicles in almost real-time in an efficient way for urban scenarios.

We propose Back off-based Per-hop Forwarding (BPF), a broadcast-based receiver-oriented protocol that uses the destination location information to select the forwarding order among the nodes receiving the packet. BFP does not require nodes to exchange periodic messages with their neighbors communicating their locations to keep a low management message overhead. It uses geographic information about the final destination node in the header of each data packet to route it in a hop-by-hop basis. It takes advantage of redundant forwarding to increase packet delivery to a destination, what is more critical in an urban scenario than in a highway, where the road topology does not represent a challenge for forwarding.

We evaluate the performance of the BPF protocol using ns-3 and a Manhattan grid topology and compare it with well-known broadcast suppression techniques. Our results show that BPF achieves significantly higher packet delivery rates at a reduced redundancy cost.
\end{abstract}


\section{Introduction}
Vehicular ad-hoc networks (VANET) were motivated mainly by safety and traffic management applications, followed by infotainment applications that provide an additional commercial utilization of the new communication infra-structure. Alternatively, we propose to use a VANET as the infrastructure for an urban cyber-physical system, an approach that has not been extensively explored so far. 

Vehicles equipped with a wide range of sensing devices and the ability to communicate with each other offer a unique opportunity for gathering real-time data about a city, like traffic conditions, environmental parameters, video and audio for surveillance~\cite{Gerla_2011}, or physical condition of the drivers~\cite{Rodrigues}. A good overview of existing work on using vehicles or VANET for sensing can be found in~\cite{Gerla_2011}. Existing VANET solutions either apply on-demand querying for local dissemination within the VANET~\cite{lee_IEEETVT2009}, sometimes keeping the data in the location it pertains to~\cite{dikaiakos_JSACA2007}, or rely on delay-tolerant networking and open Wi-Fi access points for sending the data to the Internet backbone~\cite{Hull_SenSys2006}. However, the first are inefficient for real-time traffic or environmental monitoring due to the query overhead and the need to globally access the data, and the latter cannot guarantee up-to-date data. Knowing the updated state of the various relevant variables for a city is necessary for applications such as navigation using real-time traffic information for regular and for emergency vehicles, or personal mobility and environmental monitoring. 

The purpose of sensing in the sense of a cyber-physical system is to provide the sensed data to entities outside the VANET in almost real-time. This corresponds to a system architecture where several or all nodes in the VANET are data sources and the ultimate destination of the data lies outside the VANET, whereby data can get there through one or more gateways. This article proposes and evaluates a broadcast-based protocol for data collection over VANET. 

In scenarios of high node density broadcast storms impair communication in VANET. Several algorithms have been proposed to mitigate them mostly in scenarios of safety message dissemination in highways, with some techniques focusing on reducing the amount of forwarders at each hop using probabilistic forwarding and suppression~\cite{Tonguz,Tonguz2,techniques}, some relying on using exchanged neighbor information to explicitly limit the amount of forwarders~\cite{Hartenstein}. 

We consider that it is inefficient to continuously exchange neighbor information in a high density volatile network for several reasons. First, there is the overhead of periodically exchanging the neighbor list. Second, additional mechanisms must verify whether the chosen forwarder actually forwards the packet. Third, another major reason for not using explicit choice of a single forwarder is that, in urban scenarios, this choice would require knowledge of the road topology and car density towards the destination to avoid routing packets to a dead-end. And it does not seem feasible to do routing on the road topology on a packet-by-packet basis. 

Instead, we take the approach of adding the geographic location of current forwarder and the destination to each data packet, and use this information at the receivers to rank them as potential forwarders in a distributed fashion on a packet by packet basis. The potential forwarders are differentiated using back off timers and suppression is used to limit the amount of forwarders, extending existing techniques to the urban sensing scenario. We evaluate the proposed protocol using the NS3 simulator for large scale simulation and compare its performance with well-known broadcast-based protocols in an urban setting. We analyze networking metrics, like packet delivery rate, end-to-end delay and number of hops in the path, as well as the amount of replicas that reach the destination, i.e. the redundancy added by the protocol, and the overhead in terms of total amount of packets created in the network.

The rest of the paper is organized as follows: in the next section, we present related work. Section~\ref{sec:data-gathering} describes the novel protocol and its parameters. The simulation setup is described in Section~\ref{sec:simulation}. Section~\ref{sec:performance} shows the results of the performance evaluation of the protocol, and finally Section~\ref{sec:conclusions} concludes the paper.

\section{Related Work}
\label{sec:rel-work}
There are different approaches for sending data from a node to a gateway node: one is to have data delivery routes between each node and a gateway node, created and maintained by routing protocols, as is common in mobile ad-hoc networks (MANET); another is to use hop-by-hop decisions based on the geographic location of the destination and previous forwarding nodes until the gateway is reached. 

Ad hoc On-Demand Distance Vector (AODV)~\cite{AODV} and Dynamic Source Routing (DSR)~\cite{DSR} are reactive protocols originally designed for MANETs. A number of studies have simulated and compared the performance of these protocols for VANETs~\cite{feasibility,practical}. In~\cite{feasibility}, the authors introduce prediction-based AODV protocols: Predicted AODV (PRAODV) and Predicted AODV with Maximum lifetime (PRAODVM) that uses the speed and location information of nodes to predict the link lifetime. But these methods depend on the accuracy of the prediction method, which can be low in volatile networks. Another approach is to use cluster-based protocols to improve network scalability, which create a virtual network infrastructure by clustering the nodes. Many cluster-based routing protocols~\cite{clustering}-~\cite{dominating} have been studied in MANETs. But these techniques are very unstable in VANETs and clusters created by these techniques are too short-lived. 

On the other hand, in per-hop forwarding each node, upon receiving a packet, decides to forward it. Most of existing approaches for gathering data in urban environments use the prior exchanging packets or hello messages to gain information about the neighbors or network topology around the forwarder node and then select a node which is suitable for forwarding~\cite{Contention,Knowledge,Geographic}. This prior exchanging of data has drawbacks like high network overhead, high delay and in the case of mobility does not use accurate position information. Using geographical position information in VANETs is more common and routing protocols that use this information have higher performance than topology based protocols like AODV and DSR~\cite{metropolis,Location}.

One of the well-known protocols in this category is the greedy routing protocol~\cite{Greedy} that always forwards the packet to the closest node to the destination by exchanging hello message to gain information about its neighbors. Greedy Perimeter Stateless Routing (GPSR)~\cite{GPSR} consists of two different forwarding methods: greedy forwarding and perimeter forwarding. In this method, a beaconing algorithm is used for determining the neighbor position. In~\cite{Location}, the authors showed that geographical protocol like GPSR achieves better performance compared to DSR protocol. Lochert et al.~\cite{Geographic} proposed Geographic Source Routing (GSR) which uses the city digital map to get the destination position. By combining the geographical routing and knowledge of the city map, GSR has better average delivery rate, smaller total bandwidth consumption and similar latency of first delivered packet than DSR and AODV in urban area. However, the per-packet computation overhead is very high.

DV-CAST~\cite{DV-CAST} uses sender-oriented forwarding and has three major components: neighbor detection, broadcast suppression and store-carry-forward mechanism. It uses hello messages to estimate the network topology and GPS information to determine the direction of vehicles for broadcasting the data, reducing protocol overhead and complexity. Simulation results show that DV-CAST performs well in heavy traffic during rush hours and very light traffic during certain hours of the day and also is robust against various extreme traffic conditions but still need prior hello messages.

Contention-based forwarding (CBF)~\cite{Contention} does not exchange messages to create and maintain routes between source and destination. CBF works in three steps: one hop exchanging hello messages, contention period and suppression. Upon receiving a packet, all suitable neighbors are exchanging hello messages as single hop broadcast before exchanging data packets. After that all nodes compete with each other to gain the right to forward the packet. During the contention period, each node evaluates itself to know how it is suitable to be next forwarder. This method is based on a timer. All nodes that receive packets check remaining progress to the destination and node with minimal progress will have smaller timer than other nodes and will be selected as the next forwarder. Next phase is to avoid forwarding of other nodes and establish itself as the next forwarder. For this, the selected forwarder broadcast a packet to its neighbors and each node that received this packet cancel its timer and will not forward the packet. They try to avoid having multiple forwarders by using 3 different suppression techniques: basic, area-based and active selection. In the 2 first techniques, there is a possibility of having more than one forwarder but active selection technique by using request-to-forward (RTF) and clear-to-forward (CTF) prevents all forms of packet duplications. In the CBF, each node should exchange hello messages for one hop and also to avoid other nodes to forward the packets it needs to send other hello messages. In our protocol, we avoid this kind of hello messages to reduce the network load and congestion.

Another approach is~\cite{Knowledge} which uses velocity information of nodes to predict the direction of movement. Each node has three states: \textit{Away, Still and Towards}. When nodes move in opposite direction of destination, they are in the \textit{Away} state and when they are stand, they are in the \textit{Still} state and finally nodes move toward to the destination are in the \textit{Towards} state. Only nodes with state which has been changed from \textit{Away} to \textit{Still}, \textit{Still} to \textit{Towards} or \textit{Away} to \textit{Towards} will forward the packet. In the case of similar states, if the distance of current node is smaller than distance of previous node to the destination, then it forwards the packet. This approach to gain information about the velocity and direction of other nodes needs to exchange control messages that has network overhead and reduces the data rate in the network.

Finally, we introduce in more detail 3 broadcast-based protocols that use basic per-hop forwarding and suppression techniques to mitigate broadcast storms: Weighted p-Persistence, Slotted 1-Persistence and Slotted p-Persistence broadcasting~\cite{techniques}. We shall compare the performance of the proposed protocol against these protocols because they follow a similar approach of not requiring the exchange of neighbor information. In weighted p-persistence forwarding, each node $j$, upon receiving a packet from node $i$, verifies the packet ID and re-broadcasts the packet with probability $p_{ij}$ if it receives the packets for the first time, otherwise it discards the packets. The probability of broadcasting is calculated from the distance between nodes $i$ and $j$ ($D_{ij}$) relative to the average communication range ($R$): 
\begin{equation}
\label{pij}
p_{ij} = (\frac{D_{ij}}{R})
\end{equation}

In slotted 1-persistence forwarding, each node $j$, upon receiving a packet from node $i$, checks the packet ID and re-broadcasts the packet at a timeslot $T_{S_{ij}}$ if it receives the packets for the first time and does not receive any duplicates before the assigned timeslot, otherwise it discards the packets (suppression). $T_{S_{ij}}$ is calculated by the following expression:
\begin{equation}
\label{pij}
T_{S_{ij}} = S_{ij} \times{} \tau,
\end{equation}

where $\tau$ is the estimated 1-hop delay and \textit{$S_{ij}$} is the assigned slot number, calculated by:
\begin{equation}
\label{pij}
  S_{ij} = \left\{ 
  \begin{array}{l l}
   N_s \times{} (1-\frac{D_{ij}}{R}) & \quad {D_{ij} \leq R} \\
  \\
   0 & \quad {D_{ij} > R}\\
  \end{array}, \right.
\end{equation}
and $N_s$ is the predetermined number of slots.

Finally, slotted p-persistence is a mix of the two previous approaches. Upon receiving a packet from node $i$, node $j$ checks the packet ID and re-broadcasts the packet with probability $p_{ij}$ at assigned timeslot $T_{S_{ij}}$ if it receives the packets for the first time and has not received any duplicates before assigned timeslot, otherwise it discards the packets. 
\begin{figure}[h]
\begin{center}
\caption{Node configuration used for calculations}
\label{fig:array}
\includegraphics[width=1\columnwidth]{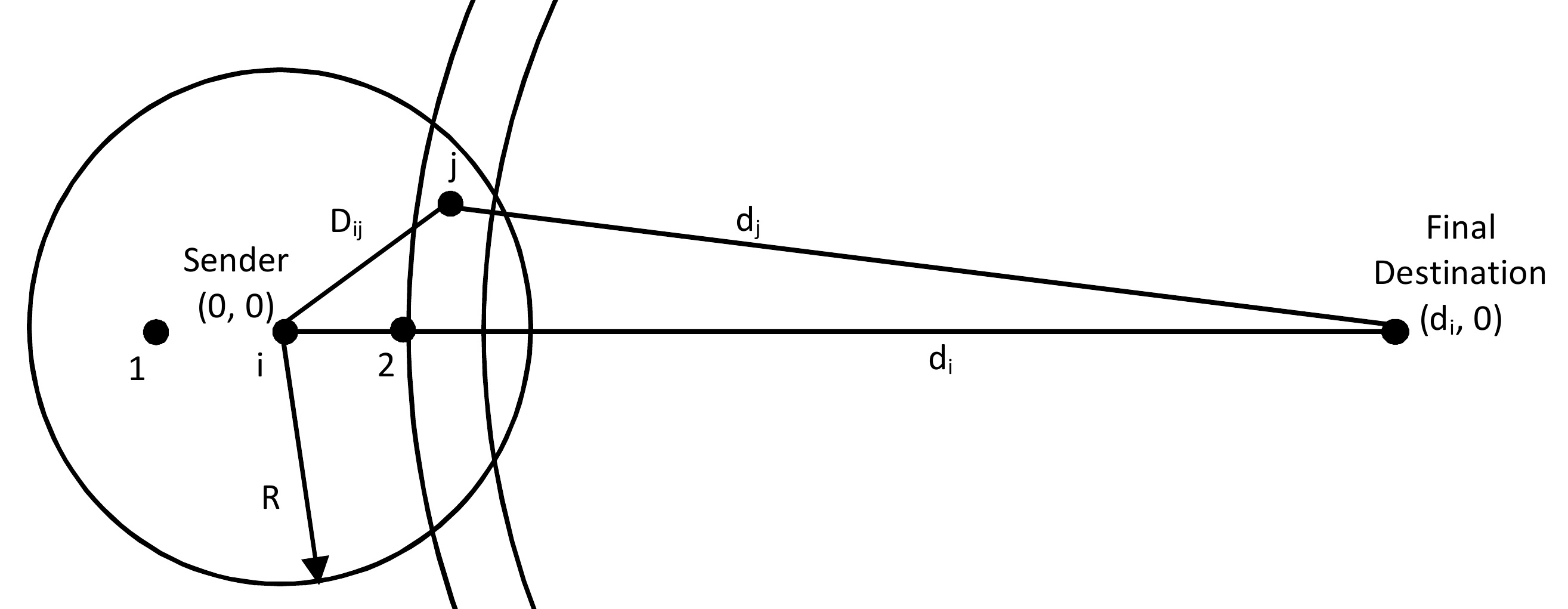} 
\end{center}
\end{figure}
\section{Broadcast-based Data Gathering Protocol}
\label{sec:data-gathering}
The protocol proposed aims at collecting large amounts of data from sensors installed in vehicles in an urban environment, configuring a cyber-physical system for an urban area. We envision that vehicles move within the urban environment and collect information like pollution or traffic conditions, and that the data generated in each node is periodically sent to a back office using the VANET as sensing infra-structure. In this scenario, we have many-to-one communication pattern from sources to the final destination and we assume that each node knows its own geographical location information and that of the final destination. The goal of the data gathering protocol is to collect this data with high packet delivery ratio (PDR), limited delay and low amount of overhead using a VANET. It is more critical in a scenario where all nodes are data sources than in other VANET scenarios to avoid congestion collapse by limiting the amount of packets forwarded in the network. 

We propose Back off-based Per-hop Forwarding (BPF), a data gathering protocol that uses the location information to select the forwarding order among the nodes receiving the packet by mapping it into back off time, so that nodes likely to be nearer to the final destination have shorter back off times. BFP has the following properties: 1) it does not require nodes to exchange periodic messages with their neighbors communicating their locations to keep low the management message overhead; 2) it uses geographic information about the current sender and the final destination node in the header of each data packet to route it in a hop-by-hop basis; 3) it takes advantage of redundant forwarding to increase packet delivery to a destination. The novelty of this protocol is the use of the final destination for per-hop forwarding in a unicast urban scenario. It takes advantage of the geographic location of the destination to direct the forwarding towards the destination, being more efficient than destination-agnostic protocols commonly used for safety message dissemination. Moreover, it takes advantage of redundancy to be more effective than protocols that specify one single forwarder, since specifying one single per-hop forwarder in an urban environment requires additional knowledge of the full street map towards the destination, or the chances are high that a packet is routed to a dead-end or along a very long route. 

\subsection{BPF Protocol Design}
Figure~\ref{fig:array} illustrates the scenario in considered to explain the calculation of the per-hop back off time: node $i$ is the previous hop for node $j$ and nodes 1, 2 and $j$ are potential forwarders.

The most straightforward choice for forwarding is the node geographically closest to the final destination~\cite{GPSR}, but that information is not available when neighbor nodes do not exchange their locations with each other. So, the preferred forwarders are the nodes that represent the most progress from the previous sender, which are the nodes located closer to the end of the transmission range, which is taken by protocols like CBF or the 1-persistent broadcasting. We further reduce the amount of forwarding nodes using the distance to the final destination in component of the back off calculation.

Usually, there are 2 constant values used to calculate the back off time: $D_{ij}$ and $d_j$. These values are distance to the last hop and to the final destination, respectively. To analyze the effect of these two constant on the back off time, we define two different components according to these values in this section: $C_1$ and $C_2$.
\begin{figure}[h]
\begin{center}
\caption{Back off time in microseconds for different positions around a node located at $(0, 0)$ according to $C_1$ and $C_2$. Location of the final destination $(2000, 0)$ (horizontally to the right of the plot) and communication range is 500~m.}
\label{fig:oldwithc}
\subfigure[$C_1$ and $C_2$]{\includegraphics[width=0.9\columnwidth]{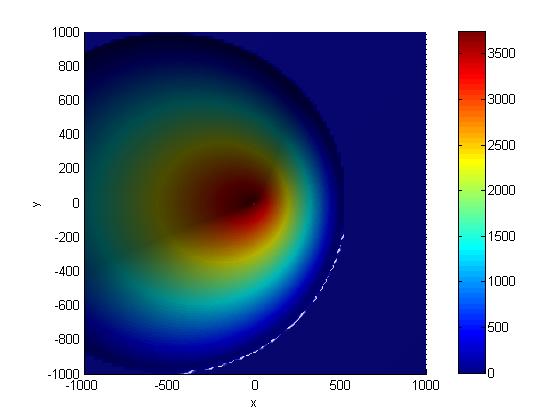}}
\subfigure[$C_2$ only]{\includegraphics[width=0.9\columnwidth]{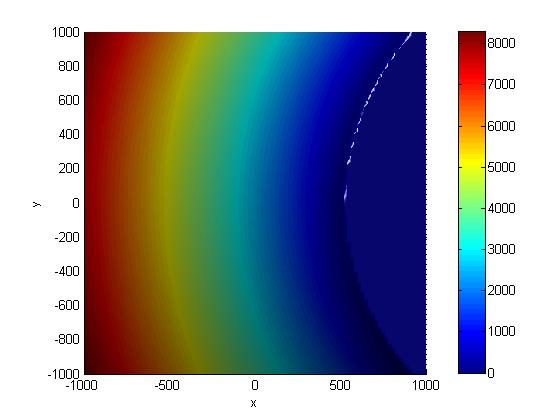}}
\end{center}
\end{figure}

The first component is the distance between current node ($j$) and previous hop ($i$), $D_{ij}$, compared with the average communication range ($R$).
By selecting nodes farther from the previous node to forward sooner than other nodes is a receiver-based greedy approach that makes packets travel the largest possible distance at each hop. We define the following component, which selects a node at the end of the communication range to have lower back off time than other nodes:
\begin{equation}
\label{c1}
C_1 = (1 - \frac{D_{ij}}{R})
\end{equation}

However, we wish to further concentrate the preferred forwarders in the direction of the final destination, since nodes at the end of the communication range in the opposite direction of the final destination can cause useless increase in the number of replicas (in Figure~\ref{fig:array} node 1 is in this situation). The second component in the calculation of the back off time is the distance to the final destination, $d_j$, relative to the distance between the previous hop and the final destination, $d_i$:
\begin{equation}
\label{c2}
C_2 = (1.0 + \frac{dj - di - R}{2R})
\end{equation}

Figure~\ref{fig:oldwithc} shows the back off time calculated as a combination of $C_1$ and $C_2$. In these plots, the previous node is located at $(0, 0)$, the final destination is at $(2000, 0)$ (horizontally to the right of the plot) and the communication range equals 500~m. The first plot depict the back off as an equally weighted sum of both components, while the second plot considers only $C_2$. 
The initial plot shows only little directionality towards the destination, because the $C_1$ component is dominant in the sum due to the fact that $D_ij$ is much larger than $d_j-d_i-R$ except when the nodes are very close to the destination. Therefore, for the evaluation of the BPF, we only consider the calculation based on the $C_2$ component.

\subsection{How to map the back off value to time?}
After calculating the back off components, we need to map this value to the back off time. Unlike other broadcast storm mitigation techniques~\cite{techniques}, we do not use the WAIT\_TIME;  we just use different back off times to forward the packet and to distribute forwarding events along the time.
The protocols in~\cite{techniques} use a WAIT\_TIME of 5000~$\mu$s to suppress as many duplicate packets as possible from previous forwarders. But we aim at forwarding the packet as fast as possible by the best positioned nodes and cancel forwarding from nodes not so well positioned to reduce the amount of transmission in the network. So we give the shortest back off time to the node with the most progress from the previous forwarder and that transmission will suppress forwarding on nodes with less progress to the destination.

In our protocol, the back off value calculated from the components is between 0 and 1, and it is multiplied by 5000~$\mu$s which is the WAIT\_TIME in the known broadcast suppression techniques~\cite{techniques}. So, the back off time at any hop lies in the interval [0,5]~ms. Note that this is the back off time of the routing protocol and the MAC layer exponential back off algorithm is run for every packet passed to the MAC layer.

\subsection{Back off-based forwarding algorithm}
\begin{figure}[h]
\begin{center}
\leavevmode
\caption{Flowchart of data gathering Protocol}
\includegraphics[width=0.9\columnwidth]{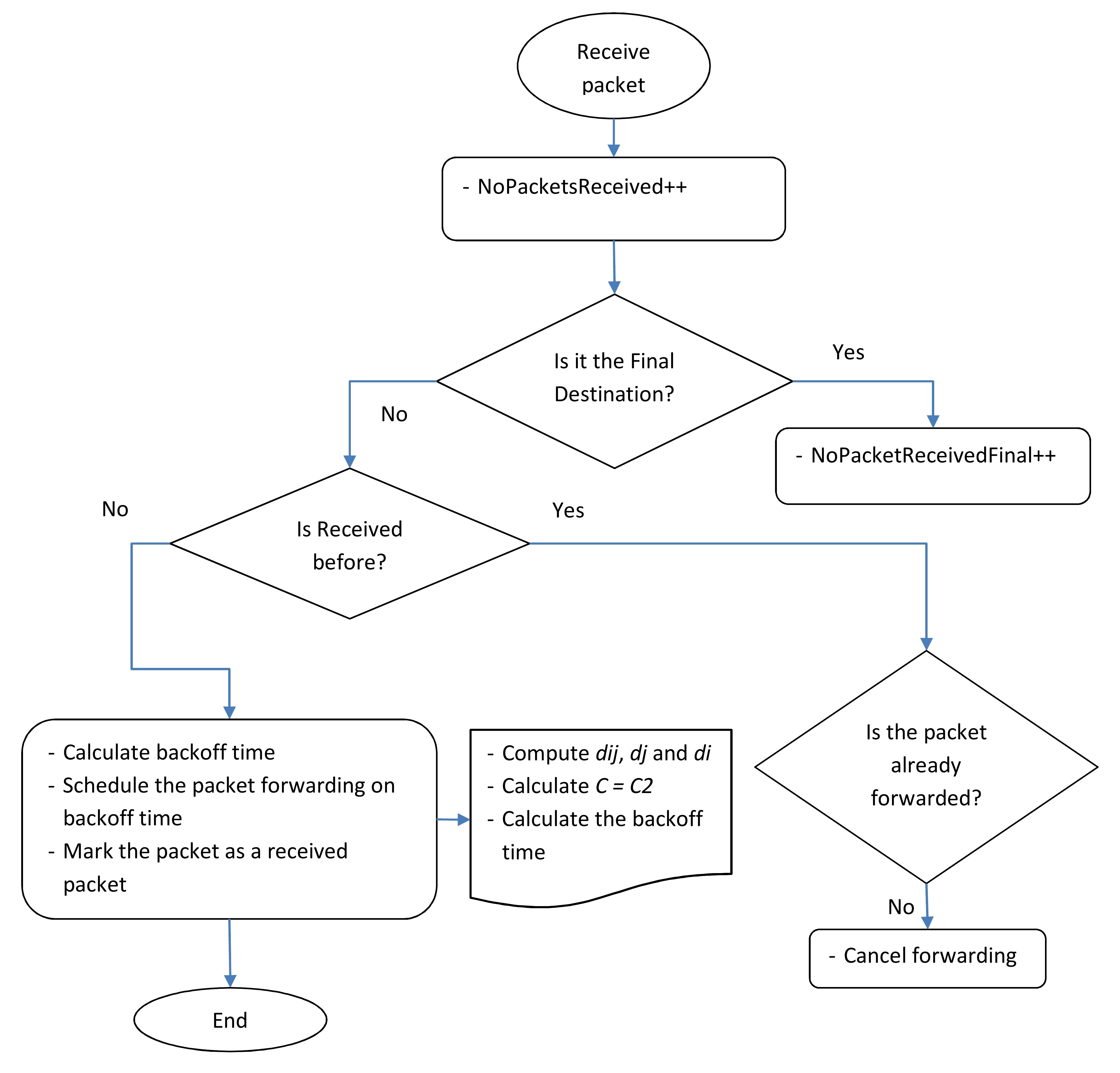}
\label{fig:main}
\end{center}
\end{figure}

The flow diagram of the per-hop forwarding algorithm executed in each node upon reception of a packet is shown in Figure~\ref{fig:main}. Each node upon receiving a packet, checks if it is the final destination node. If not, it checks if it received the packet before. If so, it cancels the forwarding event if the back off time has not expired. In the case that it receives the packet for the first time, it calculates the back off time and schedules the forwarding event on the back off time and marks the packet as a received packet.

\section{Simulation}
\label{sec:simulation}
We used the Network Simulator 3 (ns-3)~\cite{ns3} version 3.9. The topology used for movement of cars was Manhattan Grid with size $5\times5$ (from $(0m, 0m)$ to $(2500m, 2500m)$), so the simulated area was $2.5 km\times2.5 km$ with $25 km$ road length, and the final destination was located at $(1250m, 1250m)$ in the center of topology.

For the first results, and to keep feasible simulation durations, we simulate with a limited number of source nodes. We selected 8 source nodes located as far as possible in the topology. Node density is set to 2.4, 4.8, 7.2 and 9.6 nodes/km, totaling 61, 121, 181 and 241 nodes, respectively. The communication range has been set to 500~m, and each node had on average at least 2 nodes (low node density) and at most 20 nodes (high node density and at the intersections) within the communication range. Nodes move with average speed of $14 m/s$ and minimum speed of 3~m/s without pause time. Each source node sends 512~Bytes packets with rate of 5 packets/s $20 kbps$ and the simulation time is 200~seconds.

The underlying MAC protocol is set to 802.11p with PHY data rate equal to 6~Mbps and channel bandwidth is 10~MHz. The propagation loss model used in the simulation environment was Nakagami-m Propagation Loss~\cite{Nakagami,Shigehiko} with $m=1.55$ which is the recommended value for urban environments~\cite{Rubio}.

Each simulation configuration is done for 4 different protocols: BPF using only $C_2$, weighted p-persistence, slotted 1-persistence and slotted p-persistence with $p=0.5$.
For each combination of above parameters we ran 10 independent simulation runs and the results show the average and $95\%$ confidence interval for each metric.

\section{Performance Evaluation}
\label{sec:performance}
For evaluating the performance of the BPF protocol we use the following metrics: packet delivery ratio (PDR\%), end-to-end delay between source nodes and the final destination, number of hops in the path, and number of replicas of a packet that reach the final destination.

\subsection{Packet Delivery Ratio}
\begin{figure}[h]
\begin{center}
\leavevmode
\caption{PDR\% for 4 different protocols with 8 sources ($20 kbps$)}
\includegraphics[width=0.9\columnwidth]{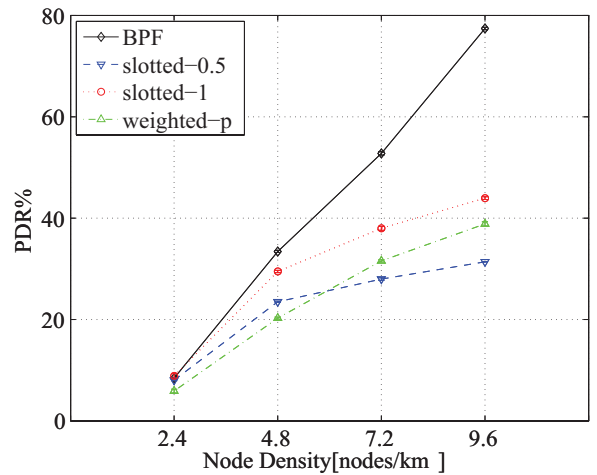} 
\label{pdr8s20}
\end{center}
\end{figure}

Figure~\ref{pdr8s20} shows the packet delivery ratio between sources and the final destination for the 4 mentioned protocols at four different traffic densities. 
BPF achieves higher end-2-end PDR\% than any of the other 3 protocols. In areas of low node density all protocols have the same PDR\% performance because there are few nodes within the communication range to forward the packet and in many cases forwarded packets end in a dead-end. As the node density increases, BPF shows increasingly better behavior than other protocols.

The BPF protocol significantly improves the end-to-end PDR\% in high node density by leveraging packet redundancy in the network. This effect overwhelms the additional collisions caused by the redundant forwarding. 

As shown in Figure~\ref{pdr8s20}, the PDR\% increases from 8\% to 78\% for the BPF protocol when the node density increases from 2.4 to 9.6~nodes/km, which is 85\% more than the PDR\% of the second best protocol in a well-connected network in high node density. Moreover, the PDR\% increasing tendency is higher than that of the other 3 protocols, which seem to start saturating at the maximum node density simulated.

\subsection{End-to-End Delay}
\begin{figure}[h]
\begin{center}
\leavevmode
\caption{Delay for 4 different protocols with 8 sources ($20 kbps$)}
\includegraphics[width=0.9\columnwidth]{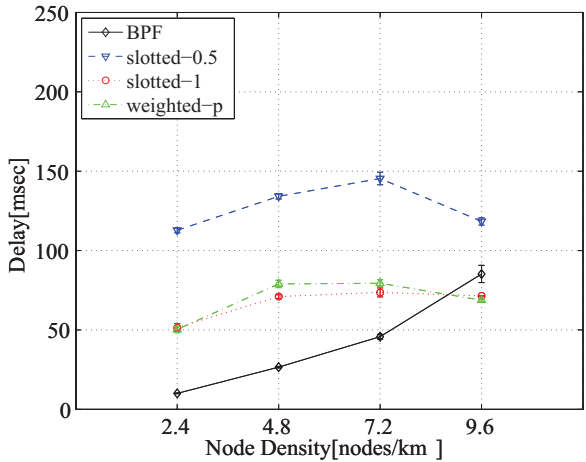} 
\label{delay4s20}
\end{center}
\end{figure}

Figure~\ref{delay4s20} shows the end-to-end delay between sources and the final destination, which is an important metric since we aim at providing up-to-date data about a city in a timely manner. BPF also has lower end-to-end delay when compared with the 3 other protocols, mainly because it does not have a WAIT\_TIME of 5~ms on each hop, as do the other protocols.

When node density increases, the number of collisions increases because there are more nodes in the communication range of any node, and the probability of having a back off time near 0 increases because of the higher number of nodes at the end of communication range. 

On the other hand, the broadcast mitigation protocols can deal better with this, because they use the WAIT\_TIME before forwarding, sender nodes receive more duplicate packets from neighbor nodes before forwarding and choose the nearest node to itself for its calculation, so the probability of collision and the delay decrease. This slump is more significant for the protocols that use the probability for forwarding (slotted-0.5 and weighted-p). 
The protocols will not forward the packets with probability $1-p$, reducing the probability of collisions, but enough other nodes forward the packets and the delay will decreases.

\subsection{Number of Hops and Amount of Replicas}
\begin{figure}[h]
\begin{center}
\leavevmode
\caption{Number of hops for 4 different protocols with 8 sources ($20 kbps$)}
\includegraphics[width=0.9\columnwidth]{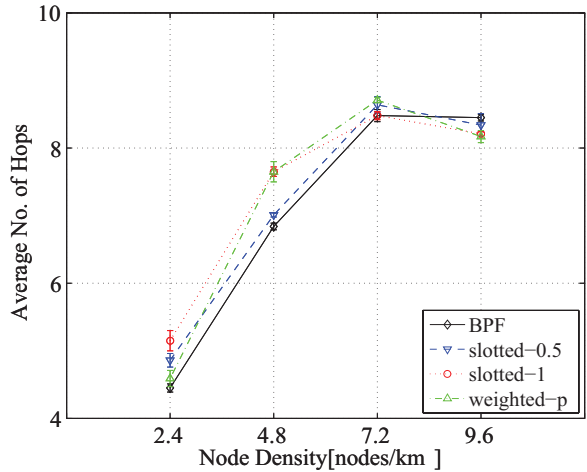} 
\label{hops4s20}
\end{center}
\end{figure}

Figure~\ref{hops4s20} shows the number of hops from source to the final destination. The number of hops is the same for all the protocols for different node density because all of the protocols try to choose the nearest node to the final destination only in different ways. 

\begin{figure}[h]
\begin{center}
\leavevmode
\caption{Average Number of replicas per uniquely received packets for 4 different protocols with 8 sources ($20 kbps$)}
\includegraphics[width=0.9\columnwidth]{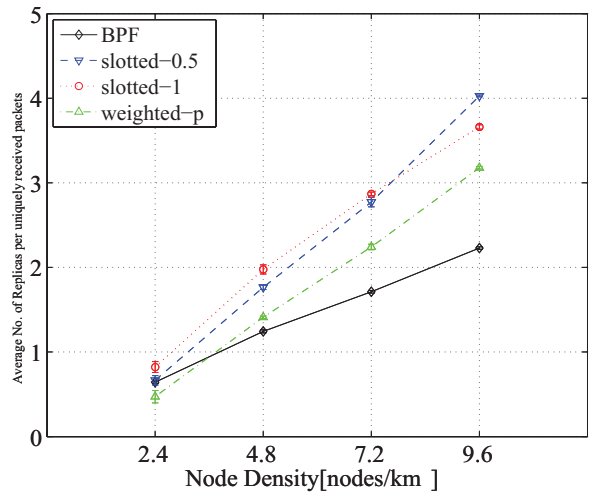} 
\label{rep20}
\end{center}
\end{figure}

As discussed before, BPF distributes the back off time and gives shorter back off time to the nodes which are nearer to the final destination, trying to reduce the number of replicas through suppression. The other protocols do it by using WAIT\_TIME and allowing for the reception of all possible packets with the same ID from neighbor nodes and then using one of those packets to calculate whether or when to forward the packet. As Figure~\ref{rep20} shows, the BPF without using WAIT\_TIME produces the same number of replicas at the final destination, showing that our protocol achieves higher PDR\% with the same redundancy as the other protocols, i.e. it is more efficient.

\subsection{Scaling Source Nodes}
\begin{figure}[h]
\begin{center}
\leavevmode
\caption{PDR\% for different number of source with 9.6 nodes/km}
\includegraphics[width=0.9\columnwidth]{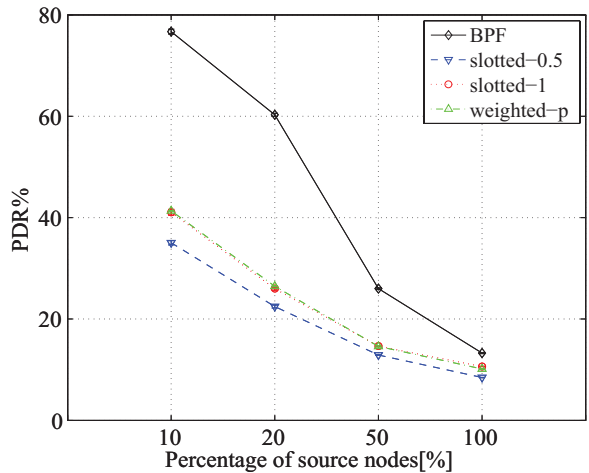} 
\label{source5}
\end{center}
\end{figure}

Since we envision a scenario where all nodes can be data sources, Figure~\ref{source5} shows the PDR\% for increasing percentage of nodes being network sources in the highest node density (9.6 nodes/km) scenario previously considered. As expected, as the number of source nodes in the network increases, the PDR\% decreases due to increasing network congestion. Nevertheless, the proposed protocol shows a higher PDR\% in all situations, showing a higher efficacy. As a future work, we will study a way to increase the PDR\% for high node density having all nodes as source nodes by decreasing the number of useless forwarding.

\section{Conclusion and Future Work}
\label{sec:conclusions}
We envision the usage of a VANET as an infra-structure for an urban cyber-physical system that makes available up-to-date data about various parameters of an urban area to services outside of the network. The main consumers of such data are applications like as traffic management or navigation using real-time traffic information for regular and for emergency vehicles, or personal mobility and environmental monitoring. The data gathering system is modeled as a many-to-one communication over VANET, a scenario that has not been previously addressed. In this scenario, broadcast storms are more likely to happen than in regular scenarios, since there are more source nodes regularly sending packets, so the amount of congestion in the network should be carefully monitored.

In this paper, we propose the Back off-based Per-hop Forwarding (BPF), a broadcast- and receiver-based per-hop forwarding protocol that selects the forwarding order among the nodes receiving the packet by mapping it into back off time, so that nodes likely to be nearer to the final destination have shorter back off times. BFP has the following properties: 1) it does not require nodes to exchange periodic messages with their neighbors communicating their locations to keep low the management message overhead; 2) it uses geographic information about the current sender and the final destination node in the header of each data packet to route it in a hop-by-hop basis; 3) it takes advantage of redundant forwarding to increase packet delivery to a destination.

We evaluated the proposed protocol and compared its performance to broadcast storm mitigation techniques for safety message dissemination using ns-3. The results show that the proposed protocol achieves higher packet delivery rates and uses on average the same number of hops and causes less redundant packets at the data sink. When subject to increasing load due to increasing number of nodes generating data, all studied protocols significantly reduce the PDR\%, although BFP maintains a higher delivery efficacy. 

However, the results also indicate that there is still room for improving the performance in higher load scenarios, which will be the focus of the next steps. Another matter of interest in this context is analyzing the effect of the length of the path between sources and destinations and limiting the amount of network congestion observed near the sink, a typical problem from sensor networks. Moreover, we will study the impact of location and number of infrastructure nodes on the data gathering capacity of the network.

\addtolength{\textheight}{-2cm}   
\bibliographystyle{IEEEtran}
\bibliography{ref}

\begin{thebibliography}{10}
\providecommand{\url}[1]{#1}
\csname url@samestyle\endcsname
\providecommand{\newblock}{\relax}
\providecommand{\bibinfo}[2]{#2}
\providecommand{\BIBentrySTDinterwordspacing}{\spaceskip=0pt\relax}
\providecommand{\BIBentryALTinterwordstretchfactor}{4}
\providecommand{\BIBentryALTinterwordspacing}{\spaceskip=\fontdimen2\font plus
\BIBentryALTinterwordstretchfactor\fontdimen3\font minus
  \fontdimen4\font\relax}
\providecommand{\BIBforeignlanguage}[2]{{%
\expandafter\ifx\csname l@#1\endcsname\relax
\typeout{** WARNING: IEEEtran.bst: No hyphenation pattern has been}%
\typeout{** loaded for the language `#1'. Using the pattern for}%
\typeout{** the default language instead.}%
\else
\language=\csname l@#1\endcsname
\fi
#2}}
\providecommand{\BIBdecl}{\relax}
\BIBdecl

\bibitem{Gerla_2011}
\BIBentryALTinterwordspacing
M.~Gerla and L.~Kleinrock, ``Vehicular networks and the future of the mobile
  internet,'' \emph{Comput. Netw.}, vol.~55, pp. 457--469, February 2011.
  [Online]. Available: \url{http://dx.doi.org/10.1016/j.comnet.2010.10.015}
\BIBentrySTDinterwordspacing

\bibitem{Rodrigues}
J.~Rodrigues, F.~Vieira, T.~Vinhoza, J.~Barros, and J.~Cunha, ``A non-intrusive
  multi-sensor system for characterizing driver behavior,'' Sep 2010, pp.
  1620--1624.

\bibitem{lee_IEEETVT2009}
U.~Lee, E.~Magistretti, M.~Gerla, P.~Bellavista, and A.~Corradi,
  ``Dissemination and harvesting of urban data using vehicular sensing
  platforms,'' \emph{IEEE Transactions on Vehicular Technology}, vol.~58,
  no.~2, pp. 882 --901, feb. 2009.

\bibitem{dikaiakos_JSACA2007}
M.~Dikaiakos, A.~Florides, T.~Nadeem, and L.~Iftode, ``Location-aware services
  over vehicular ad-hoc networks using car-to-car communication,'' \emph{IEEE
  Journal on Selected Areas in Communications}, vol.~25, no.~8, pp. 1590
  --1602, oct. 2007.

\bibitem{Hull_SenSys2006}
\BIBentryALTinterwordspacing
B.~Hull, V.~Bychkovsky, Y.~Zhang, K.~Chen, M.~Goraczko, A.~Miu, E.~Shih,
  H.~Balakrishnan, and S.~Madden, ``Cartel: a distributed mobile sensor
  computing system,'' in \emph{Proceedings of the 4th international conference
  on Embedded networked sensor systems}, ser. SenSys '06.\hskip 1em plus 0.5em
  minus 0.4em\relax New York, NY, USA: ACM, 2006, pp. 125--138. [Online].
  Available: \url{http://doi.acm.org/10.1145/1182807.1182821}
\BIBentrySTDinterwordspacing

\bibitem{Tonguz}
O.~Tonguz, N.~Wisitpongphan, F.~Bai, P.~Mudalige, and V.~Sadekar,
  ``Broadcasting in vanet,'' \emph{Mobile Networking for Vehicular
  Environments}, pp. 7--12, May 2007.

\bibitem{Tonguz2}
O.~Tonguz, N.~Wisitpongphan, J.~Parikh, F.~Bai, P.~Mudalige, and V.~Sadekar,
  ``On the broadcast storm problem in ad hoc wireless networks,'' \emph{3rd
  International Conference on Broadband Communications, Networks and Systems},
  pp. 1--11, Oct 2006.

\bibitem{techniques}
N.~Wisitpongphan, O.~Tonguz, J.~Parikh, P.~Mudalige, F.~Bai, and V.~Sadekar,
  ``Broadcast storm mitigation techniques in vehicular ad hoc networks,''
  \emph{IEEE Wireless Communications}, 2007.

\bibitem{Hartenstein}
M.~Sepulcre, J.~Gozalvez, J.~H{\"a}~andrri, and H.~Hartenstein, ``Contextual
  communications congestion control for cooperative vehicular networks,''
  \emph{Wireless Communications, IEEE Transactions on}, vol.~10, no.~2, pp. 385
  --389, Feb 2011.

\bibitem{AODV}
\BIBentryALTinterwordspacing
C.~Perkins, E.~Belding-Royer, and S.~Das, ``Ad hoc on-demand distance vector
  (aodv) routing,'' Tech. Rep., Feb 2007. [Online]. Available:
  \url{http://www.ietf.org/rfc/rfc3561.txt}
\BIBentrySTDinterwordspacing

\bibitem{DSR}
\BIBentryALTinterwordspacing
D.~Johnson, Y.~Hu, and D.~Maltz, ``The dynamic source routing protocol (dsr)
  for mobile ad hoc networks for ipv4,'' Tech. Rep., Feb 2007. [Online].
  Available: \url{http://www.ietf.org/rfc/rfc4728.txt}
\BIBentrySTDinterwordspacing

\bibitem{feasibility}
V.~Namboodiri, M.~Agarwal, and L.~Gao, ``A study on the feasibility of mobile
  gateways for vehicular ad-hoc networks,'' \emph{in Proceedings of the First
  International Workshop on Vehicular Ad Hoc Networks}, pp. 66--75, 2004.

\bibitem{practical}
S.~Wang, C.~Lin, Y.~Hwang, K.~Tao, and C.~Chou, ``A practical routing protocol
  for vehicle-formed mobile ad hoc networks on the roads,'' \emph{in
  Proceedings of the 8th IEEE International Conference on Intelligent
  Transportation Systems}, pp. 161--165, 2005.

\bibitem{clustering}
C.~Lin and M.~Gerla, ``Adaptive clustering for mobile wireless networks,''
  \emph{IEEE Journal of Selected Areas in Communications}, vol.~15, no.~7, pp.
  1265--1275, 1997.

\bibitem{dominating}
J.~Wu and H.~Li, ``A dominating-set-based routing scheme in ad hoc wireless
  networks,'' \emph{the special issue on Wireless Networks in the
  Telecommunication Systems Journa}, vol.~3, pp. 63--84, 2001.

\bibitem{Contention}
H.~Fusler, J.~Widmer, M.~Kasemann, M.~Mauve, and H.~Hartenstein,
  ``Contention-based forwarding for mobile ad hoc networks,'' \emph{Ad Hoc
  Networks}, vol.~1, no.~4, pp. 351--369, 2003.

\bibitem{Knowledge}
J.~LeBrun, C.-N. Chuah, D.~Ghosal, and M.~Zhang, ``Knowledge-based
  opportunistic forwarding in vehicular wireless ad hoc networks,''
  \emph{Vehicular Technology Conference}, vol.~4, pp. 2289--2293, May 2005.

\bibitem{Geographic}
C.~Lochert, M.~Mauve, H.~Fussler, and H.~Hartenstein, ``Geographic routing in
  city scenarios,'' \emph{ACM SIGMOBILE'05}, vol.~9, no.~1, pp. 69--72, 2005.

\bibitem{metropolis}
G.~Liu, B.-S. Lee, B.-C. Seet, C.~Foh, K.~Wong, and K.-K. Lee, ``A routing
  strategy for metropolis vehicular communications,'' \emph{in International
  Conference on Information Networking (ICOIN)}, pp. 134--143, 2004.

\bibitem{Location}
H.~F{\"u}{\ss}ler, M.~Mauve, H.~Hartenstein, M.~Kasemann, and D.~Vollmer,
  ``Location based routing for vehicular ad-hoc networks,'' \emph{ACM SIGMOBILE
  Mobile Computing and Communications Review (MC2R)}, vol.~7, no.~1, pp.
  47--49, Jan 2003.

\bibitem{Greedy}
J.~Bernsen and D.~Manivannan, ``Greedy routing protocols for vehicular ad hoc
  networks,'' \emph{Wireless Communications and Mobile Computing Conference},
  pp. 632--637, Aug 2008.

\bibitem{GPSR}
B.~Karp and H.~Kung, ``Gpsr: Greedy perimeter stateless routing for wireless
  networks,'' \emph{in Proceedings of the ACM/IEEE International Conference on
  Mobile Computing and Networking (MobiCom)}, 2000.

\bibitem{DV-CAST}
O.~Tonguz, N.~Wisitpongphan, and F.~Bai, ``Dv-cast: A distributed vehicular
  broadcast protocol for vehicular ad-hoc networks,'' \emph{IEEE Wireless
  Communications}, 2010.

\bibitem{ns3}
\BIBentryALTinterwordspacing
``Network simulator 3.'' [Online]. Available: \url{http://www.nsnam.org/}
\BIBentrySTDinterwordspacing

\bibitem{Nakagami}
``ns3::nakagamipropagationlossmodel class reference,'' July 13, 2011.

\bibitem{Shigehiko}
O.~Shigehiko, ``Nakagami-m fading channel,'' \emph{Journal of the Institute of
  Electronics, Information and Communication Engineers}, vol.~86, no.~12, pp.
  969--971, 2003.

\bibitem{Rubio}
L.~Rubio, J.~Reig, and N.~Cardona, ``Evaluation of nakagami fading behaviour
  based on measurements in urban scenarios,'' \emph{International Journal of
  Electronics and Communications}, vol.~61, pp. 135--138, Feb 2007.

\end{thebibliography}
\end{document}